\documentclass[aps,preprint,superscriptaddress,showpacs]{revtex4}
\usepackage[english]{babel}
\usepackage[dvips]{graphicx}
\usepackage{amsmath}
\usepackage{amsfonts}
\usepackage{amssymb}
\usepackage[linktocpage,breaklinks=true,colorlinks=true,linkcolor=black,filecolor=black,citecolor=black,pagecolor=black,urlcolor=black]{hyperref}

\usepackage{color}

\usepackage{epstopdf}

\newcommand{\cf}{\emph{cf.\ }}

\usepackage{graphicx}
\begin{document}

\title{Collective scattering and oscillation modes of optically bound point particles trapped in a single mode waveguide field}
\date{\today}

\author{Daniela Holzmann}
\email{Daniela.Holzmann@uibk.ac.at}

\author{Helmut Ritsch}
\affiliation{Institute for Theoretical Physics, University of Innsbruck, Technikerstra\ss e 25, A-6020 Innsbruck, Austria}

\begin{abstract}
Collective coherent scattering of laser light induces strong light forces between polarizable point particles. These dipole forces are strongly enhanced in magnitude and distance within the field of an optical waveguide so that at low temperature the particles self-order in strongly bound regular patterns. The stationary configurations typically exhibit super-radiant scattering with strong particle and light confinement. Here we study collective excitations of such self-consistent crystalline particle-light structures as function of particle number and pump strength. Multiple scattering and absorption modify the collective particle-field eigenfrequencies and create eigenmodes of surprisingly complex nature. For larger arrays this often leads to dynamical instabilities and disintegration of the structures even if additional damping is present.\end{abstract}


\maketitle

\section{Introduction}

Laser light scattering off free particles is accompanied by momentum transfer and thus forces. If the incident light fields are far detuned from any optical resonance, energy absorption and spontaneous re-emission play only a minor role and a coherent redistribution of photon momenta determines the force. As such coherent scattering from different particles largely preserves the laser phase, these fields interfere and thus the resulting force on each particle depends on its distance to nearby particles and acquires a collective nature, which depends on all other particle positions~\cite{birkl1995bragg,black2003observation}. While in a 3D geometry the interaction strongly decays with distance and is averaged out by phase randomization from particle positions and motion, in regular particle arrays or in a spatially confined geometry, this interaction can become very strong~\cite{piovella2001superradiant, demergis2012ultrastrong}. Particularly strong effects appear if the light fields are optically confined in resonators~\cite{ritsch2013cold}, guided by optical structures as mirrors~\cite{labeyrie2014optomechanical} or confined in waveguides~\cite{chang2013self,griesser2013light,holzmann2014self,shahmoon2014giant}. Besides cold dilute gases, closely related effects have been studied using various kinds of nano-particles in solution~\cite{dholakia2010colloquium,singer2003self,fournier2004building}.

Following first predictions by Chang and coworkers for near resonant weak pumping of atomic dipoles~\cite{chang2013self,shahmoon2014giant}, we have recently shown that cold particles illuminated far of resonance, who can freely move along or within an optical nano-fiber field, tend to form regular but complex optical structures through optical long range coupling. They collectively scatter light into the fiber mode and self-trap in the optical potential generated by interference of the scattered light and the pump field~\cite{holzmann2014self}. Surprisingly we see that not only the light confines the particle motion but the particles also confine the light forming Bragg like atom mirrors at their outer edges. Numerical simulations predict that in certain cases even more complex structures as cavity arrays are self-formed by the particles~\cite{griesser2013light}. This potentially generates a self-ordered cavity QED system with strong particle photon coupling via so called atomic mirrors~\cite{chang2012cavity}. While such dipole energy minimizing configurations were also studied for conventional 1D optical lattices, they could be shown to be generally unstable~\cite{asboth2007comment}.

In this work we theoretically study the dynamics and stability of such self-organized cavity QED systems. Already in our previous work we could identify configurations where all forces on the particles vanish and one gets a restoring force towards these equilibrium positions for small shifts for each particle. Nevertheless, displacing one particle immediately changes the fields at all other particle positions and creates perturbation forces on them. Here we cannot expect any energy conservation for the particle motion as we are dealing with an open system where the pump laser forms a non depleting energy and momentum reservoir~\cite{asboth2007collective,ostermann2014scattering}. Hence perturbing one particle induces nonconservative collective motion of all particles. Eventually these oscillations exhibit exponential growth and lead to a disintegration of the whole structure. A central aim of this work is to calculate and study the frequencies and stability of the corresponding linearized eigenmodes of the coupled atom field systems and their application to induce tailored long range interactions in this system. Of course the system is inherently nonlinear so that the full dynamics can only be captured numerically.  

Our work is organized as follows. First we quickly overview the results of our recent paper~\cite{holzmann2014self}. Out of these, in order to investigate the stability of the particles, we linearize the forces acting on the particles around their equilibrium positions. Calculating the eigenvalues of the so obtained coupling matrix we find a condition which enables and limits the formation of stable configurations. With this we first study configurations in the negligible coupling limit and then pass over to more realistic complex coupling parameters.

\section{Scattering matrix approach to light induced motion in 1D}

\begin{figure}
 \centering
 \includegraphics[width=.75\columnwidth]{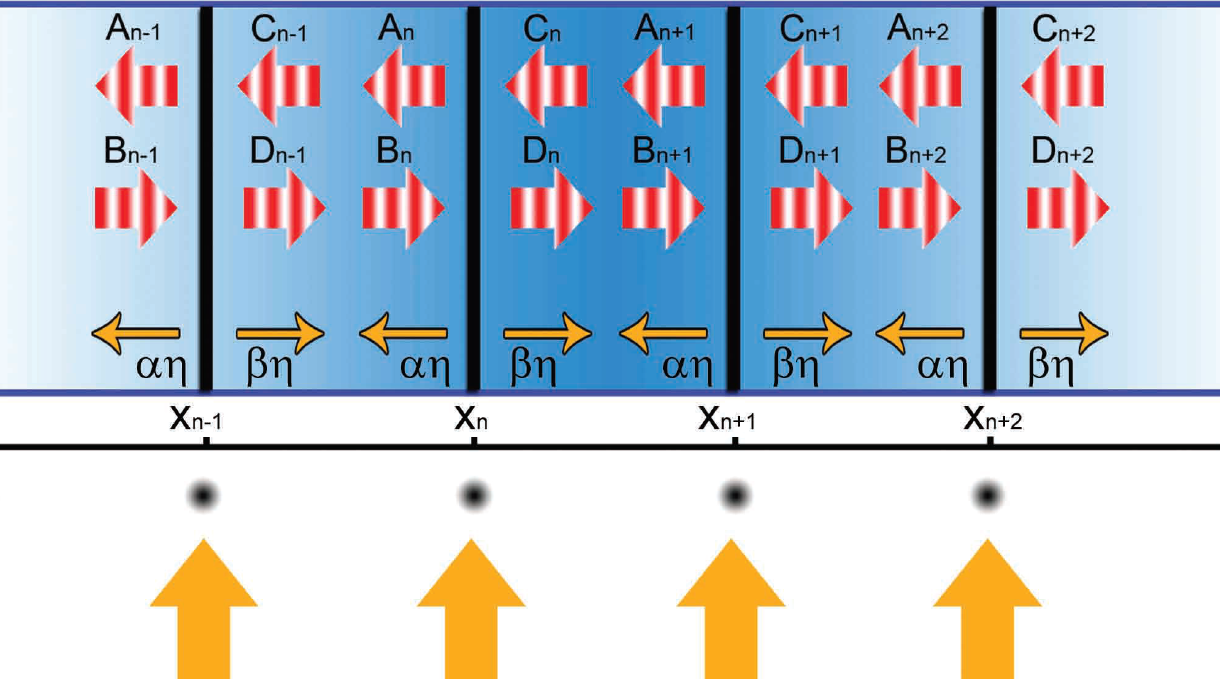}
 \caption{A 1D array of point particles scattering light in and out of an optical nano-structure can be modelled as a collection of beam splitters interacting with a plane wave.}
\label{fig_system}
\end{figure}

Let us first briefly recall our previously developed multiple scattering approach to calculate the stationary fields and forces on a linear array of particles with transverse illumination as schematically depicted in Fig.~\ref{fig_system}~\cite{holzmann2014self} . For a given spatial arrangement of point scatterers in the field of a 1D waveguide the light field can be calculated by a scattering matrix approach~\cite{asboth2008optomechanical, sonnleitner2012optomechanical,ostermann2014scattering,xuereb2009scattering}. The contribution of each particle is represented by a single $3 \times 3$ matrix parametrized by an effective coupling constant $\zeta$ proportional to its linear polarizability $\tilde{\alpha}$ and the field mode amplitude at its position. This determines the interaction between the particle and the fiber mode as well as the scattering amplitude in and out of the fiber  ${\eta}$. The transmission and reflection coefficients are related to the coupling constant via $t=1/(1-i\zeta)$ and $r=i\zeta/(1-i\zeta)$. For symmetric scattering the coupling between the amplitudes left and right of a particle can then be expressed using the following beam splitter matrix $\mathbf{M}_{BS}$~\cite{holzmann2014self}

\begin{equation}
	\begin{pmatrix} A_j\\B_j\\ \eta \end{pmatrix} 
		= \frac{1}{t}\begin{pmatrix} t^2-r^2 & r & \frac{1}{\sqrt{2}}( t-r) \\ -r & 1 & -\frac{1}{\sqrt{2}} \\ 0 & 0 & t  \end{pmatrix}
				\begin{pmatrix} C_j\\D_j\\ \eta \end{pmatrix} = \begin{pmatrix} 1+i\zeta & i\zeta & \frac{1}{\sqrt{2}}(1-i\zeta) \\ -i\zeta & 1-i\zeta & \frac{1}{\sqrt{2}}(i\zeta-1) \\ 0 & 0 & 1  \end{pmatrix} 
				\begin{pmatrix} C_j\\D_j\\ \eta \end{pmatrix} 
,
\end{equation}
where $B_j,C_j,\eta$ are the incoming fields allowing to determine the outgoing fields $A_j,D_j$, \cf Fig.~\ref{fig_system}. These in turn are used as inputs to neighbouring scatterers as $C_{j-1}=A_j \exp(i k (x_j-x_{j-1}))$ and $D_{j-1}=B_j \exp(-i k (x_j-x_{j-1}))$. Multiplying all the matrices including the propagation phase shifts then allows to determine the field distribution corresponding to the momentary particle positions. Note that we neglect the propagation time of the light within the structure compared to particle motion.  

Using simple arguments based on the Maxwell stress tensor the momentary field then determines the optical force on the $j$-th particle along the $x$-axis~\cite{asboth2008optomechanical,antonoyiannakis1999electromagnetic}:

\begin{equation}\label{eq_force_BSj_general}
	F_j=\frac{\epsilon_0}{2}\left(\vert A_j\vert^2+\vert B_j\vert^2-\vert C_j\vert^2-\vert D_j\vert^2\right).
\end{equation}
We will use this optical force then as ingredient to obtain Newton's equations for the particle motion, where we add damping describing additional light or background gas induced friction. In previous work we and others have seen that under quite general conditions stationary equilibrium positions can be found, where all particles are force free and feel a restoring force against small shifts. However, in contrast to a standard 1D optical lattice, the absence of a force does not enforce an unaltered propagation of the light~\cite{asboth2008optomechanical} but just guarantees a balance between internal and external light scattering~\cite{holzmann2014self}.
Following Ref.~\cite{asboth2008optomechanical} and including a linear friction term proportional to $\mu$ the equations of motion for the point particles of mass $m$ then read:

\begin{equation}
m\ddot{x}_j=-\mu \dot{x}_j+F_j(x_1,\dots,x_N).
\end{equation}
Stationary structures require $F_j(x_1,\dots,x_N) =0 $ for all $j$. Here these are not simple equidistant lattices, but rather complex particle arrangements with regular central parts including the formation of Bragg like reflectors at both ends of the equilibrium particle distribution denoted as $x_j^{(0)}$~\cite{holzmann2014self}.   

\section{Collective excitations around equilibrium positions}

When a particle is weakly displaced from its equilibrium $x_j(t) =x_j^{(0)}+\xi_j(t) $, it feels a restoring force and at the same time induces a global light field modification generating forces on all other particles. Once set free the perturbed particles start a correlated oscillatory motion around their equilibrium positions $x_j^{(0)}$. In the following we study the spatial properties and eigenfrequencies of such small amplitude particle oscillations analogous to phonons of a lattice. Note that the existence of a stationary stable equilibrium configuration does not guarantee stability with respect to even very small such collective oscillations, which in general can grow in amplitude and disintegrate the particle array. 

A Taylor expansion of the forces gives the following set of coupled differential equations

\begin{equation}
m\ddot{\xi}_j=-\mu \dot{\xi}_j+\sum_{l=1}^{N}D_{jl}\xi_l,
\label{newton}
\end{equation}
with nonlocal coupling matrix

\begin{equation}
D_{jl}=\frac{\partial}{\partial x_l}F_j(x_1=x_1^{(0)},\dots,x_N=x_N^{(0)})=\lim_{\xi\rightarrow 0}\frac{1}{\xi} F_j(x_v=x_v^{(0)}+\delta_{lv}\xi,~v=1,\dots, N).
\label{Djl}
\end{equation}
Note that these coupling constants would be an ideal basis for a coupled oscillator model in the ultracold gas limit. For the classical point particle model here we simply use the ansatz $\vec{\xi}=\vec{A}e^{i\omega t}$ in equations~\eqref{newton} and solve for the eigensystem of $\mathbf{D}$, where the oscillation frequencies can be calculated from the eigenvalues $\lambda_\nu$ via
\begin{equation}
\omega_\nu=\frac{i\mu\pm\sqrt{-\mu^2-4m\lambda_\nu}}{2m},
\label{freqencymu}
\end{equation}
yielding the linearized solutions:
\begin{equation}
\xi_j=\left(A_{j} e^{\frac{i\sqrt{-\mu^2-4m\lambda_\nu}}{2m}t}+B_{j} e^{\frac{-i\sqrt{-\mu^2-4m\lambda_\nu}}{2m}t}\right)e^{-\frac{\mu t}{2m}}.
\label{xj}
\end{equation}
The eigenfrequencies include a real part describing damping or antidamping depending on its sign. If positive, it leads to an amplification of the oscillation amplitude and the particles can not form a stable stationary configuration. Hence, to ensure stable configurations the eigenvalues have to fulfil the condition~\cite{asboth2008optomechanical}:
\begin{equation}
m (\Im(\lambda_\nu))^2\le-\mu^2 \Re(\lambda_\nu).
\label{stabcond}
\end{equation}
Note that added damping can restore the stability only as long as all real parts of the eigenvalues are negative. Without external friction, $\mu=0$, Eq.~\eqref{xj} simplifies to

\begin{equation}
\xi_j=A_{j} e^{i\sqrt{\frac{-\lambda_\nu}{m}}}+B_{j} e^{-i\sqrt{\frac{-\lambda_\nu}{m}}}
\label{xd0}
\end{equation}
and as long as we have no imaginary parts of the eigenvalues $\lambda_{\nu}$ and the real parts are negative the particles are simply oscillating around their equilibrium position.

\subsection{Negligible mode coupling limit $\zeta=0$} 

In the special case when particle mode coupling is very weak but we have a strong transverse pump field applied, one can set $\zeta=0$ while still keeping a nonzero scattering into the fiber, thus $\eta\neq 0$. Here the scattering of the field by the particles within the fiber is neglected and any two particles interact equivalently.  As predicted before~\cite{chang2013self} and also found by us in Ref.~\cite{holzmann2014self}, in this limit the particles in a stationary state are equidistantly distributed and the transfer matrix can easily be calculated explicitly:
\begin{equation}
\mathbf{M}_{TM}=(\mathbf{M}_{BS}\cdot \mathbf{P}(d))^{N-1} \cdot \mathbf{M}_{BS}=\mathbf{M}^{N-1} \cdot \mathbf{M}_{BS}.
\end{equation}
Here the collectively scattered field intensities are symmetric to right and left and read 

\begin{equation}
	I_{ol}=I_{or}=\frac{I_\eta}{2}\left(\frac{\sin(\frac{N k d}{2})}{\sin(\frac{k d}{2})}\right)^2
\end{equation}
and for the force on the $j$-th of $N$ particles we get:

\begin{equation}
	F_j=-\frac{P_\eta  \cos\left(N k d/2\right)\sin\left((2 j-N-1)k d/2\right)}{\sin\left(k d/2\right)},
\end{equation}
with $P_\eta=I_\eta/c$ the radiation pressure resulting from the transverse pump field.\\
This force vanishes for different lattice constants $ d= (2n-1)\lambda /(2N)$ with $ n\in\mathbb{N}$ giving potential stationary solutions. We will now calculate the linearized coupling matrices $\mathbf{D}$ for a small displacement of the $l$-th particle within such a solution. 
For the amplitude perturbation at the $j$-th cloud from displacing the $l$-th cloud by $\xi=\epsilon k$ we have to distinguish between the two cases if it is at the right or left side of the displaced $l$-th cloud:

\begin{equation}
\begin{pmatrix}
A_j\\B_j\\ \eta
\end{pmatrix}
=\mathbf{M}^{N-j+1}\mathbf{P}(-d)\left(\begin{pmatrix}
0\\D_N\\ \eta
\end{pmatrix}+\epsilon\begin{pmatrix}
0\\b\\0
\end{pmatrix}\right),
\end{equation}
with the perturbation $\vec{b}$. For $j<l$ the amplitudes can be calculated as follows

\begin{equation}
\begin{pmatrix}
A_j\\B_j\\ \eta
\end{pmatrix}=\mathbf{M}^{l-j}\mathbf{P}(\frac{\epsilon}{k})\mathbf{M}_{BS}\mathbf{P}(-\frac{\epsilon}{k})\mathbf{P}(d)\mathbf{M}^{N-l}\mathbf{P}(-d)\left(\begin{pmatrix}
0\\D_N\\ \eta
\end{pmatrix}+\epsilon\begin{pmatrix}
0\\b\\0
\end{pmatrix}\right).
\end{equation}

Let us first look at the stability of the solution with the maximal possible lattice distance $n=N$, i.e. $d=(2N-1)\lambda/(2N) $, where the force on the $j$-th particle induced by the perturbation of the $l$-th particle reads:

\begin{equation}
F_j\left(d=\frac{2N-1}{2 N}\lambda \right)=
\begin{cases}
P_\eta\epsilon\sin\left(\frac{\vert j-l\vert\pi}{N}\right), &\text{for}~ j\neq l\\
-P_\eta\epsilon\frac{\cos\left(\frac{\pi}{2N}\right)}{\sin\left(\frac{\pi}{2N}\right)}, &\text{for}~ j=l,
\end{cases}
\end{equation}
and so $D_{jl}=\frac{k}{\epsilon}F_j$ is:

\begin{equation}
D_{jl}=P_\eta k\left(\sin\left(\frac{\vert j-l\vert\pi}{N}\right)-\delta_{jl}\cot\left(\frac{\pi}{2N}\right)\right).
\end{equation}
$\mathbf{D}$ is a symmetric matrix with all eigenvalues $\lambda_\nu$ real and as it is a circulant matrix, they and the eigenvectors $\vec{z}_\nu$ can easily be calculated to give:
\begin{equation}
\begin{split}
\lambda_\nu
=-2 P_\eta k\frac{\cot\left(\frac{\pi}{2N}\right)\sin^2\left(\frac{\pi(\nu-1)}{N}\right)}{\cos\left(\frac{\pi}{N}\right)-\cos\left(\frac{2\pi(\nu-1)}{N}\right)}, \;\vec{z}_\nu=\left(e^{\frac{2\pi i(\nu-1)(j-1)}{N}}\right)_{j=1}^{N}.
\end{split}
\end{equation}

\begin{figure}
\centering
\includegraphics[width=0.7\textwidth]{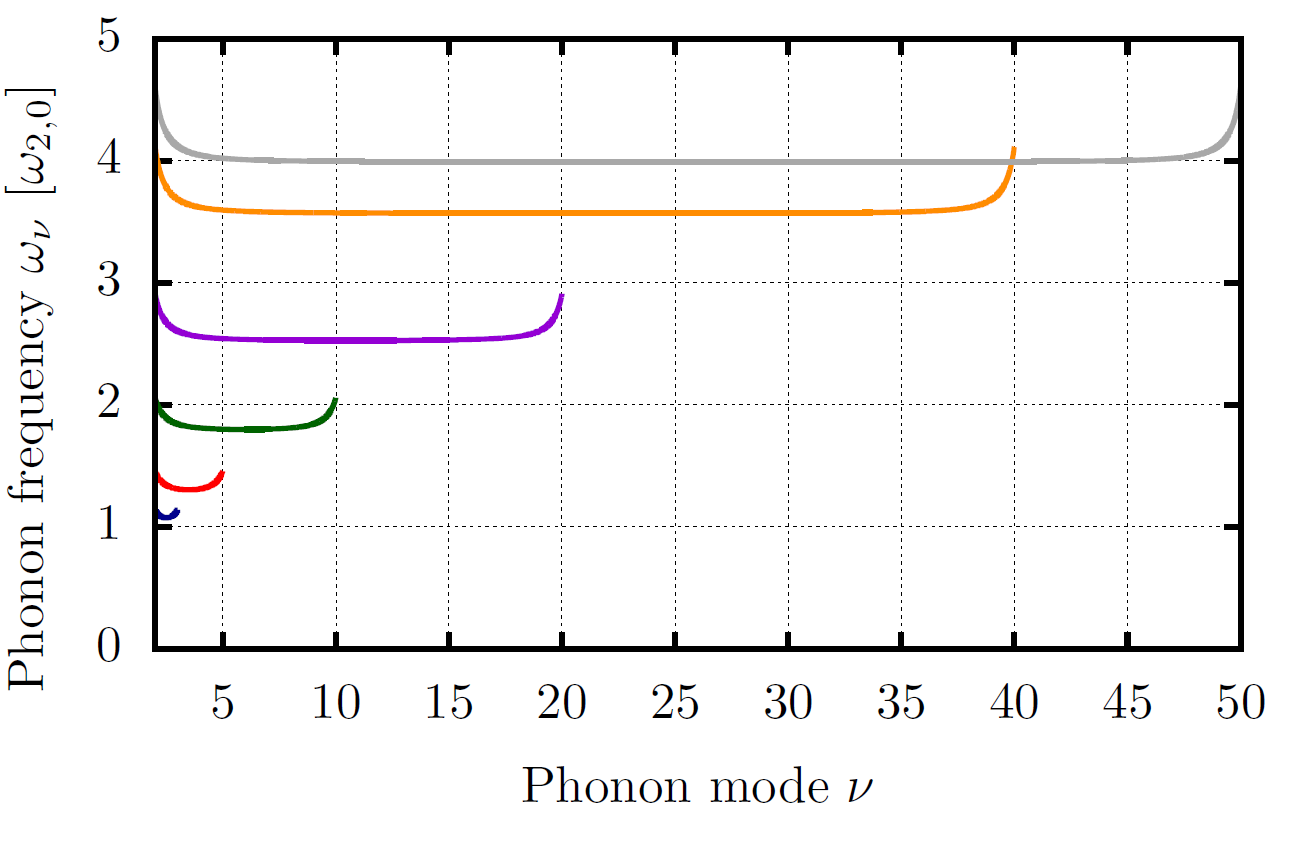} 
\caption{Phonon frequency as a function of the phonon mode $\nu$ for $\zeta=0$ beginning with $\nu=2$. The blue line corresponds to $N=3$, the red line to $N=5$, the green line to $N=10$, the violet line to $N=20$, the orange line to $N=40$ and the grey line to $N=50$.}
\label{frequency}
\end{figure}

As we have seen in Eq.~\eqref{xd0}, stably oscillating configurations require real and negative eigenvalues. Hence this configuration is stable and we find $N$ phonon modes with frequencies

\begin{equation}
\begin{split}
\omega_\nu=\sqrt{\frac{P_\eta k}{m}\left(\cot\left(\frac{\pi}{2N}\right)+\frac{\sin\left(\frac{\pi}{N}\right)}{\cos\left(\frac{\pi}{N}\right)-\cos\left(\frac{2\pi(\nu-1)}{N}\right)}\right)}
=\omega_{2,0}\sqrt{\frac{\cot\left(\frac{\pi}{2N}\right)\sin^2\left(\frac{\pi(\nu-1)}{N}\right)}{\cos\left(\frac{\pi}{N}\right)-\cos\left(\frac{2\pi(\nu-1)}{N}\right)}},
\end{split}
\end{equation}
where we have defined $\omega_{2,0}=\sqrt{\frac{2P_\eta k}{m}}$ as binding oscillation frequency for two particles at $\zeta=0$. As expected from translation invariance of the system, the lowest $\nu=1$-mode corresponding to the center of mass motion has a zero eigenfrequency. Interestingly our result exactly coincides with the prediction based on nonlocal fiber enhanced resonant dipole-dipole interaction obtained before in Ref.~\cite{chang2013self}. In the limit of large particle numbers we find:

\begin{equation}
\omega_\nu\approx 2(\nu-1)\omega_{2,0}\sqrt{\frac{N}{\pi(3+4\nu(\nu-2))}}.
\end{equation}
Consequently the particle binding frequency grows with $\sqrt{N}$ for large particle numbers stiffening the structure with particle number. Fig.~\ref{frequency} shows the phonon frequency as a function of mode number for different particle numbers. It demonstrates that the second and the $N$-th mode posses higher frequencies than all the other modes which are almost degenerate.
 
It is now interesting to see how the atoms interact with respect to their distance. For this we insert Eq.~\eqref{Djl} into Eq.~\eqref{newton} and calculate the perturbation induced forces explicitly:

\begin{align}
F_j
=P_\eta k\left(\sum_{l\neq j}\sin\left(\frac{\vert j-l
\vert\pi}{N}\right)(\xi_l-\xi_j)\right),
\end{align}
exhibiting a coupling between the oscillators proportional to $\sin\left(\frac{\vert j-l\vert\pi}{N}\right)$ depending on the distance $\vert j-l\vert$ and particle number $N$. Maximum coupling occurs between center and boundary particles at $N=2\vert j-l\vert$ and it vanishes for $\vert j-l\vert\ll N$.

\begin{figure}
\centering
\includegraphics[width=\textwidth]{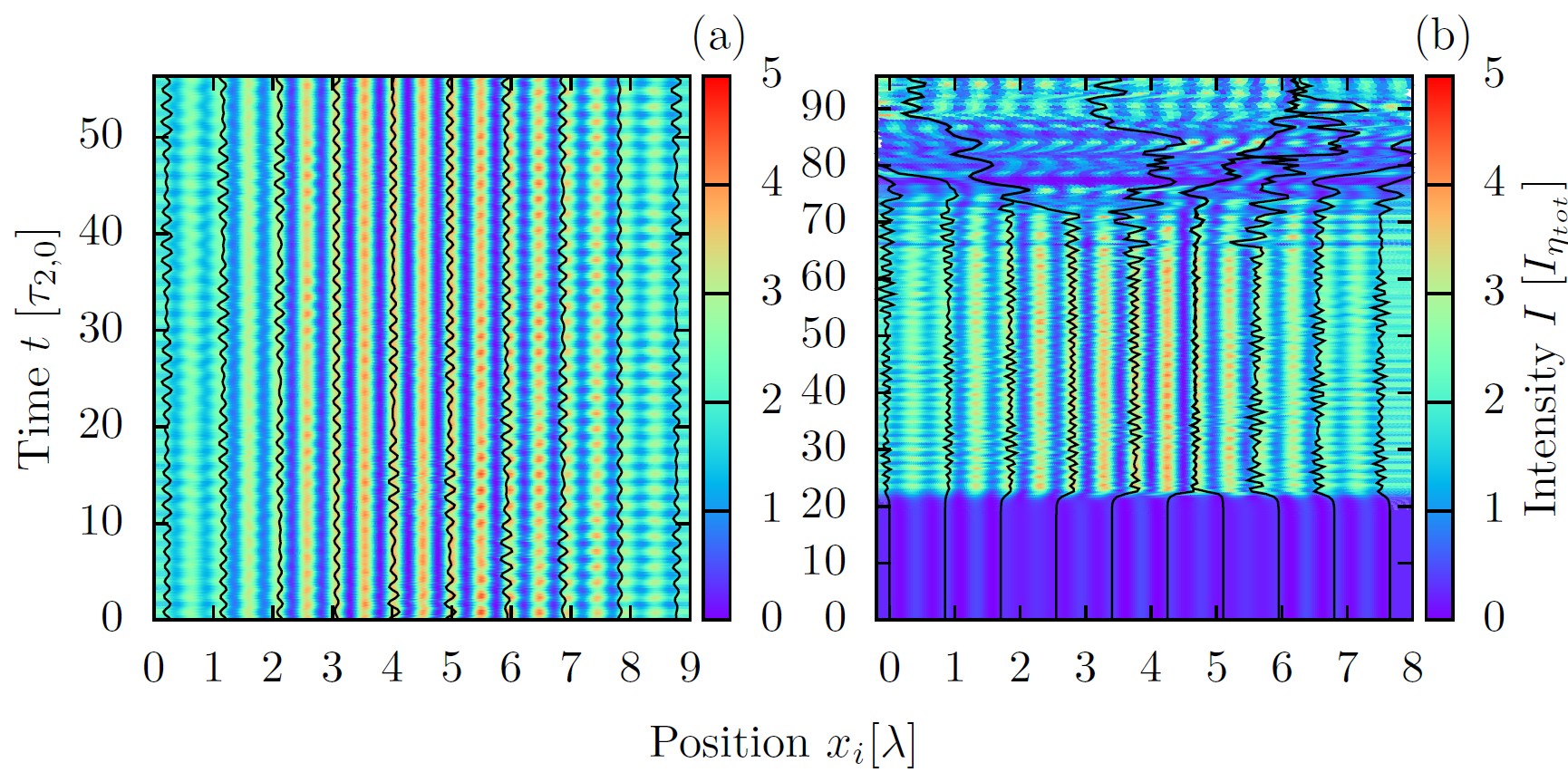}
\caption{Ten particles evolving over time with only transverse pump and $\zeta=0$. In Fig. (a) we have chosen $d=\frac{2N-1}{2N}\lambda$ ($n=N$) as initial condition slightly perturbed by $\xi_{initial}=0.1 \lambda$, while in Fig. (b) the particles are initially positioned at an unstable zero force distance $d=\frac{N-1}{N}\lambda$ ($n=N-1$). ${\tau_{2,0}}$ is defined as $\tau_{2,0}=\frac{2\pi}{\omega_{2,0}}$.}
\label{int10}
\end{figure}

Let us now insert other zero force solutions at smaller distances $d_n=\frac{2n-1}{2N}\lambda$ into Eq.~\eqref{Djl}. Here we can explicitly calculate the eigenvalues in the limit of large particle numbers

\begin{equation}
\lambda_\nu=P_\eta k \frac{2(2n-1)^2}{4(\nu-1)^2-(2n-1)^2}.
\end{equation}
As according to Eq.~\eqref{zetaimu} fully stable configurations are only possible if all eigenvalues are negative, we see that the first case, $n=N$, was the only  stable configuration with a distance below $\lambda$. An example how the particles evolve over time is plotted in Fig.~\ref{int10}, where we compare stable configurations setting $n=N$ (Fig. (a)) with unstable ones setting $n=N-1$ (Fig. (b)).
Note that starting from a stable configuration we obtain correlated oscillations of particles and fields around the stable position with a slow nonlocal excitation transfer between the sites. At negligible coupling, $\zeta =0$, one finds neither damping nor amplification of the oscillation modes but we still see a higher light intensity confined at the center of the structure. Starting from an unstable zero-force configuration the particles keep their position only for a short time until two particles collapse and the other particles reorder. Nevertheless, also this new configuration is not stable and the whole order disintegrates.

\subsection{Collective dynamics for finite particle field interaction parameter $\zeta$ }

Particles within the fiber mode will not only couple pump light in and out of the fiber, but also scatter it to the opposite propagation direction, into free space or absorb it. This is effectively described by a non-zero complex value of $\zeta$~\cite{asboth2008optomechanical}. As a first consequence this immediately leads to spatially inhomogeneous fields in the mode and thus non-equidistant stationary configurations, which, in general, do not allow for analytic treatment. Although the equation for the force~\eqref{eq_force_BSj_general} still looks very simple, its explicit form gets already complicated for more than three particles and even the zero-force points can not be found analytically. Hence to still get some analytical insights we first study systems of only two and three particles.

\subsubsection{Two particles}

\begin{figure}
\centering
\includegraphics[width=\textwidth]{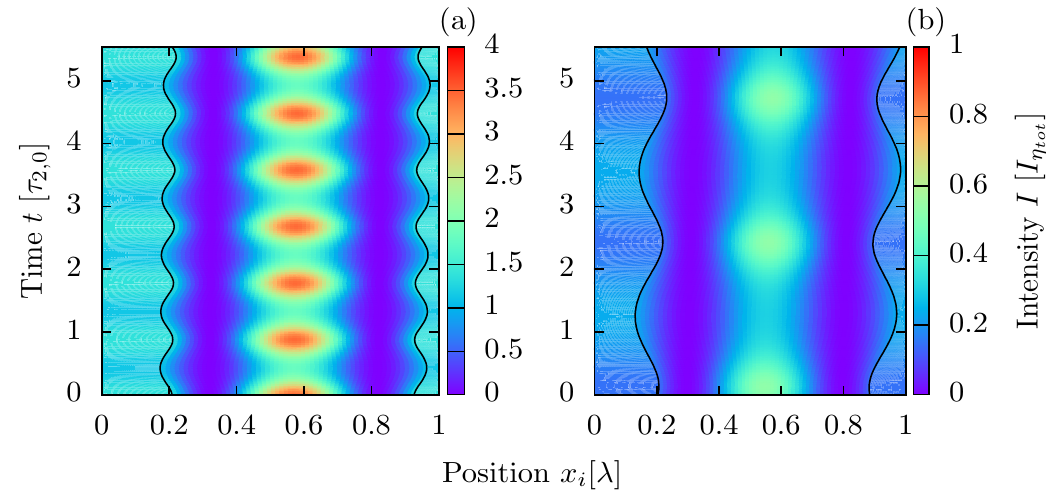} 
\caption{Two particles evolving over time with only transverse pump oscillating after a small perturbation with maximal frequency at $\zeta=0.618$ (a), and minimal frequency at $\zeta=-1.618$ (b).}
\label{int2}
\end{figure}

For two particles with no fields injected along the fiber, the light scattered by one particle only interferes with the light scattered by the second particle. Translation and spatial inversion invariance of the system here implies a vanishing sum of the forces on the two particles as long as we neglect asymmetric chiral coupling~\cite{ramos2014quantum}. The force as function of the real part $\zeta_r$ and imaginary part $\zeta_i$ then depends only on particle distance $d$~\cite{holzmann2014self}:

\begin{equation}
	F_1=-F_2=\frac{P_\eta \vert 1-i\zeta\vert^2 \cos (k d)}{4 \left(\vert\zeta \vert^2+\zeta_i\right) \cos^2(\frac{k  d}{2})+ 2 \zeta_r \sin (k  d)+1},
\end{equation}
and the zero force distances $d_0=\frac{\lambda}{2}\left(\frac{1}{2}+n\right), n\in\mathbb{N} $ are independent on $\zeta$. Interestingly even for strongly absorbing particles as e.g. gold nano-particles $\zeta_i >> \zeta_r$, we thus obtain the same force free binding distances~\cite{demergis2012ultrastrong}. Intuitively this can be understood from the $\pi/2$ relative phase shift of the two scattered fields at this distance preventing interference. To study the stability and strength of binding we calculate the coupling matrix $\mathbf{D}$: 
\begin{equation}
D_{jl}=(-1)^{l-j}\frac{P_\eta k\vert 1-i\zeta\vert^2 \left(\sin(kd)\left(2\vert \zeta\vert^2+2\zeta_i+1\right)+2\zeta_r\right)}{\left(2\left( \cos(kd)+1 \right)\left(\vert\zeta\vert^2+\zeta_i\right)+2\sin(kd)\zeta_r+1\right)^2},
\label{Djl2}
\end{equation}
and its eigenvalues. Inserting the zero force distances $d=\left(\frac{1}{4}+n\right)\lambda$ the nonzero eigenvalue reads
\begin{equation}
\lambda_2 =\frac{2k P_\eta\vert 1-i\zeta\vert^2}{1+2(\vert\zeta\vert^2+\zeta_i-\zeta_r)},
\end{equation}
and for $d=\left(\frac{3}{4}+n\right)\lambda$ we get the same eigenvalue with opposite sign. As the non-zero eigenvalue for the latter case is real and negative, these are stable configurations. The phonon frequency characterizing the optical binding strength then reads:

\begin{figure}
\centering
\includegraphics[width=\textwidth]{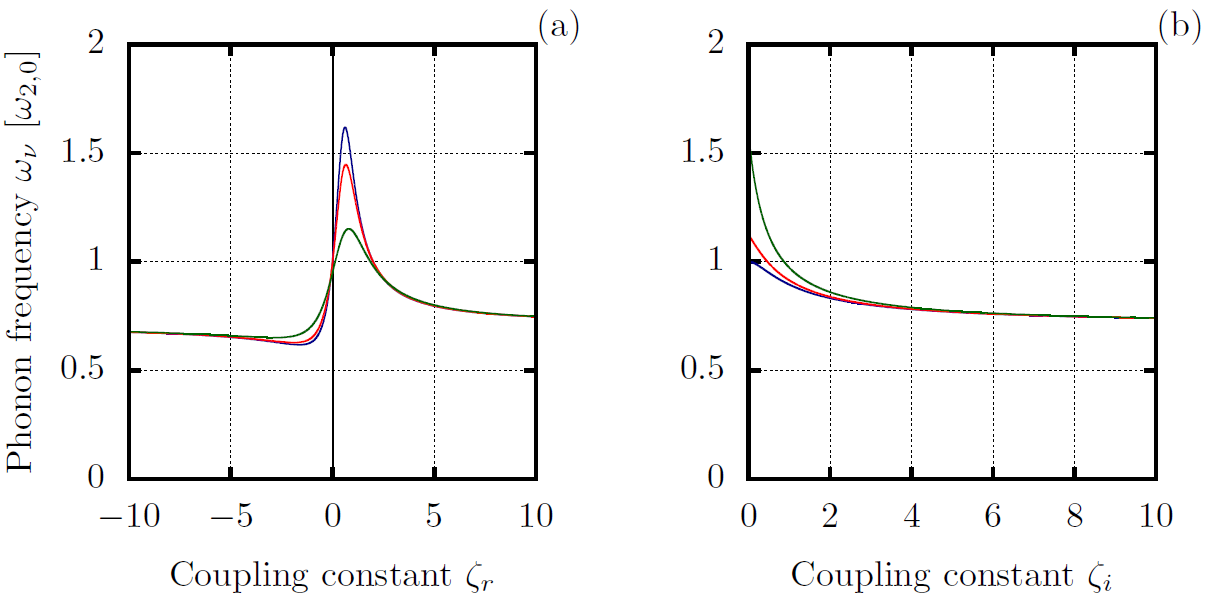} 
\caption{Fig. (a) shows the phonon frequency $\omega_\nu$ as a function of the real part of the coupling constant $\zeta_r$. The blue line corresponds to $\zeta_i=0$, the red line to $\zeta_i=\frac{1}{9}$ and the green line to $\zeta_i=\frac{1}{2}$.\\
Fig. (b) shows the phonon frequency $\omega_\nu$ as a function of the imaginary part of the coupling constant $\zeta_i$. The blue line corresponds to $\zeta_r=0$, the red line to $\zeta_r=\frac{1}{9}$ and the green line to $\zeta_r=\frac{1}{2}$.\\
We can see that the function goes to $\frac{\omega_{2,0}}{\sqrt{2}}$ for large values of $\zeta$. For the real part of $\zeta$ we can find a maximum for $\zeta_r>0$.}
\label{frequency2}
\end{figure}

\begin{equation}
\begin{split}
\omega=\omega_{2,0}\sqrt{\frac{\vert 1-i\zeta\vert^2}{1+2(\vert\zeta\vert^2+\zeta_i-\zeta_r)}},
\end{split}
\end{equation}

which for large values of $\zeta$ stays finite and converges to $\frac{\omega_{2,0}}{\sqrt{2}}$ below the noninteracting value. Nevertheless for real $\zeta$ $\omega_2$ reaches a maximum of $\omega_{2,0}\sqrt{3+\sqrt{5}}$ at $\zeta\simeq 0.618$ (Fig.~\ref{int2}(a)), while we find a minimum coupling frequency for blue detuning at $\zeta\simeq -1.618$ of $\omega_2=\omega_{2,0}\sqrt{3-\sqrt{5}}$ (Fig.~\ref{int2}(b)). The dynamics in these two extreme cases is shown in figure~\ref{int2}. Note that in both cases we can observe that the particles and fields oscillate and are periodically pushed apart whenever the intensity of the light trapped between them is maximal. Note that even in the blue detuned case for low field seeking particles we find stable trapping but much slower oscillations, i.e. weaker confinement. 


\subsubsection{Three particle dynamics}

Let us now add a third particle. Due to symmetry we will restrict ourselves to equidistant configurations and calculate the total transfer matrix. Again the amplitudes of the electric field to the left and right of the particles yield the forces acting on the particles, where as a consequence of symmetry the force on the middle particle is zero and the remaining two sum up to zero, i.e.:

\begin{equation}
\begin{split}
  F_1&=-F_3=\simeq P_\eta\left( \cos(k d)+\cos(2 k d)-\zeta  (2 \sin(2 k d)+\sin(3 k d)+\sin(4 k d))\right) + O[\zeta ]^2,\\
  F_2&=0
  \end{split}
\end{equation}

\begin{figure}
\centering
\includegraphics[width=\textwidth]{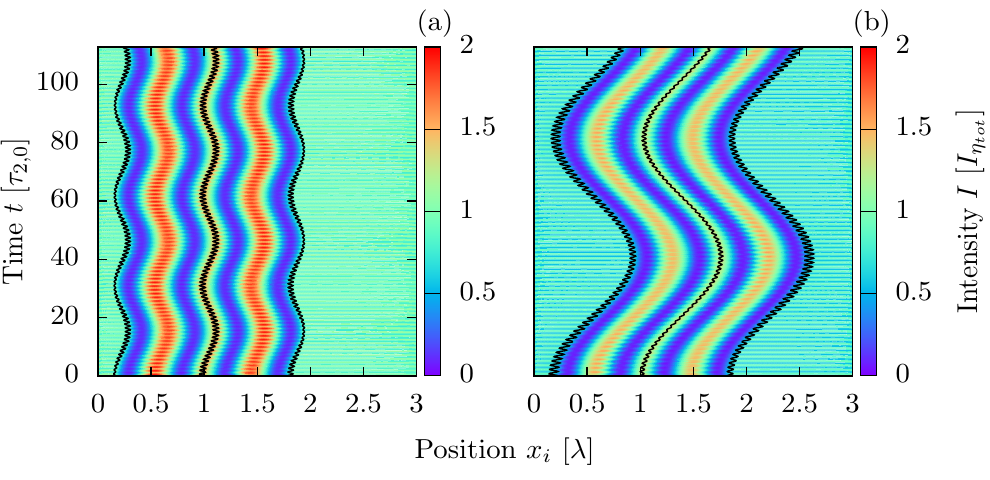} 
\caption{Three particles evolving over time with only transverse pump, $\zeta=\frac{1}{9}$ (a), $\zeta=\frac{1+i}{9}$ (b) and a small perturbation $\xi_{initial}=0.1\lambda$.}
\label{int3}
\end{figure}

 Here the general expressions are complex so that we only present some special cases to exhibit the stability of the particles. For real $\zeta=\frac{1}{9}$ only one stable equidistant configuration exists with $d_1=d_2\simeq (0.8276+n)\lambda$. Adding an imaginary part, $\zeta=\frac{1+i}{9}$, the stationary distances are enlarged to $d_1=d_2\simeq (0.8305+n)\lambda$. As shown in Fig.~\ref{int3} the trajectories of the particles here show surprising features.

Although the stationary state is symmetric with respect to the middle particle, a small perturbation from this equilibrium induces an intricate superposition of two oscillations, namely fast relative oscillations superimposed on a much slower and large amplitude a center of mass oscillation. As the system possesses translational invariance implying momentum conservation, it could be expected in principle to not allow any center of mass oscillations. This is at first sight also confirmed by the oscillation eigenvalues $\{0, -3.83 P_\eta k, -3.51 P_\eta k\}$ calculated for real $\zeta=1/9$  with eigenvectors 

\begin{equation}
z_1=\frac{1}{\sqrt{3}}\begin{pmatrix}
1\\ 1\\ 1
\end{pmatrix},~
z_2=\begin{pmatrix}
0.287\\ -0.914\\0.287
\end{pmatrix},~
z_3=\frac{1}{\sqrt{2}}\begin{pmatrix}
-1\\ 0\\ 1
\end{pmatrix}.
\end{equation} 
The eigenvalue $\lambda_1$ corresponding to the center of mass oscillation $z_1$ is indeed zero. However, we can see that the eigenvector $z_2$, corresponding to the case where two outer particles oscillate together against the middle particle includes a center of mass oscillation, as the middle particle moves much more then the sum of the outer two. This is also clearly visible in figure~\ref{int3}. At this point we have to recall that we have a strongly coupled atom field system where the pump laser constitutes an external energy and momentum reservoir, which can provide or accept momentum from the particles and conservation is only true for the combined atom field system. The asymmetric oscillation $z_2$ thus periodically channels momentum in and out of the light field inducing these unexpected center of mass oscillations. 
The dynamics gets even more complicated if we allow for light absorption and investigate the eigensystem for an imaginary coupling constant, $\zeta=\frac{1+i}{9}$. We find the eigenvalues $\{0, -2.98 P_\eta k, -2.89 P_\eta k\}$ with eigenvectors

\begin{equation}
z_1=\frac{1}{\sqrt{3}}\begin{pmatrix}
1\\ 1\\ 1
\end{pmatrix},~z_2=\begin{pmatrix}
0.349\\ -0.869\\0.349
\end{pmatrix},~
z_3=\frac{1}{\sqrt{2}}\begin{pmatrix}
-1\\ 0\\ 1
\end{pmatrix}.
\end{equation}
Again we observe a large amplitude center of mass oscillation due to the non inversion symmetric initial perturbations which lead to even amplified center of mass motion. Hence we see that while we have stationary equilibrium configuration, the system tends to disintegrate due to dynamical instability when no external friction force is provided.

\section{Numerical studies of the dynamics of larger ensembles}

\begin{figure}
\centering
\includegraphics[width=\textwidth]{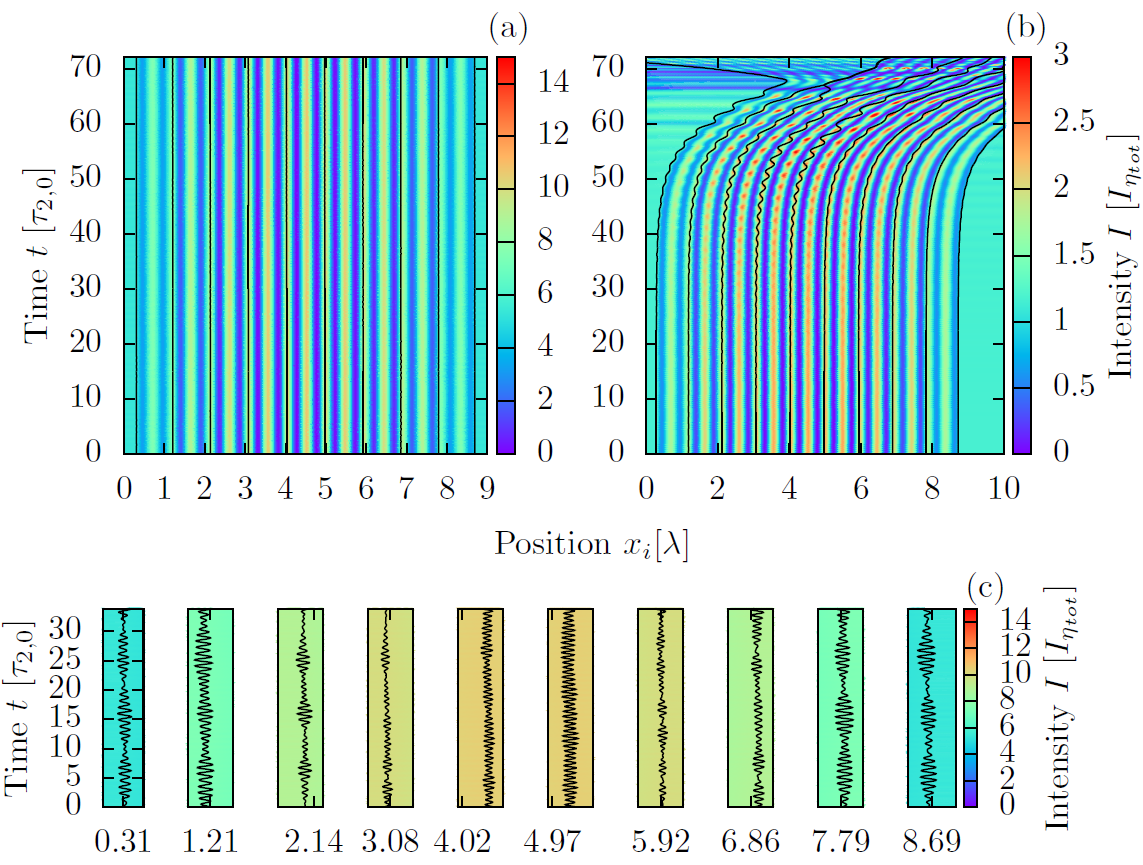} 
\caption{Ten particles evolving over time with only transverse pump, $\zeta=\frac{1}{9}$ (a), $\zeta=\frac{1+i}{9}$ (b) and a small perturbation of $\zeta_{initial}=0.001\lambda$. Fig. (c) shows the oscillation of the particles. Here we magnify the trajectories of Fig. (a) and plotted a range of $0.01\lambda$. In the second case the particles, first, start to oscillate around their equilibrium positions, but the perturbation leads to antidamping, which, in the end, destroys the configuration. The eigenvalues of the first case are all negative and real, while the eigenvalues of the second case acquire positive and negative imaginary parts.}
\label{intz10}
\end{figure}

While the momentary light field for a given configuration of many particles can still be found by simple matrix multiplication, the explicit expression for the forces soon becomes prohibitively complicated and their common zeros cannot be analytically obtained any more. Numerically, however, this still can be done efficiently even for hundreds of particles and the solutions of the corresponding dynamical equations of motion are still possible. In order to capture typical features of the resulting many particle dynamics but still obtain readable plots, here we study the case of ten particles for various coupling strengths and damping. Fig.~\ref{intz10} compares the time evolution for real polarizability $\zeta$ (Fig.~\ref{intz10}(a)) with the case of partial absorption, where $\zeta$ also possesses a small imaginary part (Fig.~\ref{intz10}(b)). In the first case a small perturbation simply induces small correlated oscillations in the vicinity of local field maxima with some slow energy exchange between the particles. In the second case with absorption, the particles start to oscillate first, but these oscillations grow until one reaches the nonlinear regime leading to a complete destruction of the order. As shown in a recent paper~\cite{holzmann2014self} one can still find a stable configuration for this case in the over damped limit. The mathematical origin for this instability is that due to the imaginary part of $\zeta$ the eigenvalues also acquire imaginary parts. This instability still can be suppressed up to a limiting value by introducing an additional friction coefficient as in Eq.~\eqref{stabcond}.

We had demonstrated that in the weak-coupling-limit, $\zeta=0$, one has stable oscillatory solutions without the need of damping. Introducing $\zeta\neq 0$ this behaviour is often lost and even for rather small perturbations the particle oscillations grow beyond the linearized regime. This leads to the collapse of the particle structure and they are expelled to the sides. It is now interesting to study the grade of instability in such a system, which turn out to be very parameter and size sensitive. To quantitatively show this we compute the eigenvalues of $\mathbf{D}$ and examine the real parts of them characterizing the magnitude of the corresponding instabilities.
\begin{figure}
\centering
\includegraphics[width=\textwidth]{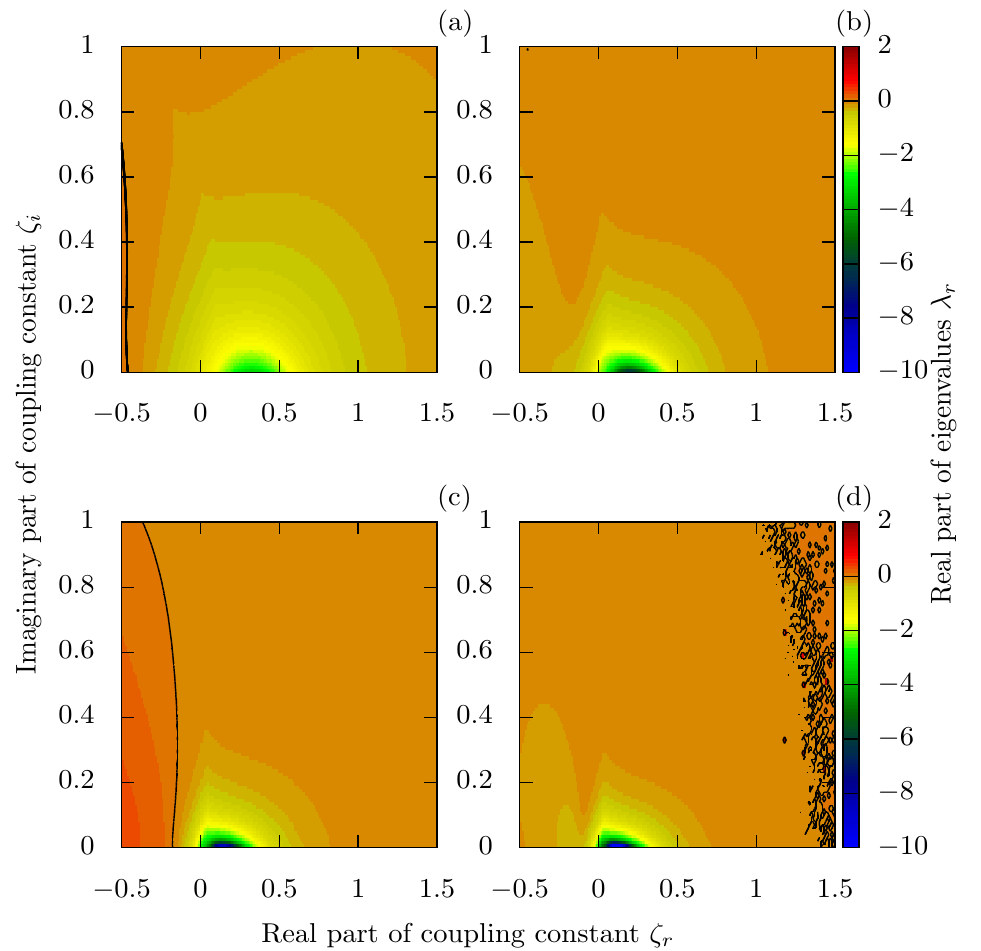} 
\caption{Maximum real part of the nonzero eigenvalues as function of real and imaginary part of $\zeta$. Fig.~(a) corresponds to $N=5$ particles, (b) to $N=10$, (c) to $N=15$ and (d) to $N=20$. Black lines show the zero value contour lines.
}
\label{zetaimu}
\end{figure}

The maximum real part of all the nonzero eigenvalues of $\mathbf{D}$ is plotted in Fig.~\ref{zetaimu} as a function of real and imaginary part of $\zeta$ for different system sizes $N$. If this value is negative, in principle one can always add enough extra damping to stabilize the configuration, while a positive value implies dynamical instability in any case. The figures show that for odd particle number we get a boundary line to the left of which at least one eigenvalue is positive and Eq.~\eqref{stabcond} thus predicts an instability, which can not be removed by damping. For negative real $\zeta$ and even particle numbers (Fig. (b) and (d)) we find stable configurations with negative eigenvalues but they are very very close to zero. In this case the system is very sensitive to small perturbations and needs strong extra damping for stabilization. Here in numerical simulations we often find unstable dynamics despite using a nominally stable parameter set, as validity of the linearized equations is very limited. 

An analogous phenomenon occurs for large particle numbers (Fig. (d)) and large values of $\zeta_r$. In this case it is difficult to distinguish between stable configurations with eigenvalues close to zero and unstable configurations. We can observe that for large $\zeta$ positive eigenvalues appear, especially for imaginary $\zeta$, which prevent the formation of stable configurations. In general only for small and mostly real $\zeta$ (blue region in Fig.~\ref{zetaimu} ) robust stable configurations can be expected. 

\section{Conclusions}
Coherently illuminated particles in a 1D trap order in strongly bound regular arrays induced by the interference of the scattered and pump light. If the scattered light is guided by nano-optical waveguide structures one finds very strong binding and long range order with phonon frequencies growing approximately with the square root of the particle number. In general the scatterers arrange in configurations, where the scattered light is resonantly confined within the structure to maximize the optical trapping potential for the particles. This can be seen as a controllable prototype 1D implementation of an atom light crystal. In this regime absorption and scattering of the fields by the particles within the waveguide create a strongly nonlinear dynamical response of the system, which limits the size of stable arrays even in the presence of additional friction forces. This hints at the possibility to create self-ordered nano-cavity QED systems where the particles simultaneously form the resonator via atom mirrors~\cite{hetet2011single} as well as the optical active system. As a next task towards a quantized field treatment one of course would need to estimate the magnitude of the self-ordered single photon coupling strength, which requires a sort of dynamic mode description. Generalizations to several frequencies, polarizations or higher order transverse modes should generate even more intriguing complex physics. In principle similar mechanisms should also be at work in planar 2D setups~\cite{fournier2004building} leading to even stronger transverse light confinement and limiting the extension of optical matter.    

{\bf Acknowledgements:}{We thank D. Chang for helpful discussions. This work has been supported by the Austrian Science Fund (FWF) through SFB Foqus project F4013.}

\bibliography{scatt1d}
\end{document}